\documentclass[12pt,twocolumn,tighten]{aastex62}
\usepackage{graphics}
\usepackage{color}
\usepackage{amsmath}
\usepackage{empheq}

\newcommand{\msun}{M$_{\odot}$}

\newcommand{\HII}{H{\sc ii}}

\shortauthors{McQuinn et al.}
\shorttitle{Using the TRGB as a Distance Indicator in the Near-IR}

\begin{document}
\title{Using the Tip of the Red Giant Branch as a Distance Indicator in the Near Infrared}

\author{Kristen.~B.~W. McQuinn}
\affiliation{Rutgers University, Department of Physics and Astronomy, 136 Frelinghuysen Road, Piscataway, NJ 08854, USA} 
\affiliation{University of Texas at Austin, McDonald Observatory, 2515 Speedway, Stop C1400 Austin, Texas 78712, USA}\email{kristen.mcquinn@rutgers.edu}

\author{Martha Boyer}
\affiliation{Space Telescope Science Institute, 3700 San Martin Drive, Baltimore, MD 21218, USA}

\author{Evan D. Skillman},
\affiliation{University of Minnesota, Minnesota Institute for Astrophysics, School of Physics and Astronomy, 116 Church Street, S.E., Minneapolis, MN 55455, USA} 

\author{Andrew E.~Dolphin}
\affiliation{Raytheon Company, 1151 E. Hermans Road, Tucson, AZ 85756, USA}
\affiliation{University of Arizona, Steward Observatory, 933 North Cherry Avenue, Tucson, AZ 85721, USA}

\begin{abstract}
The tip of the red giant branch (TRGB) is a well-established standard candle used to measure distances to nearby galaxies. The TRGB luminosity is typically measured in the I-band, where the luminosity has little dependency on stellar age or stellar metallicity. As the TRGB is brighter at wavelengths redder than the I-band, observational gains can be made if the TRGB luminosity can be robustly calibrated at longer wavelengths. This is of particular interest given the infrared capabilities that will be available with the James Webb Space Telescope and an important calibration consideration for using TRGB distances as part of an independent measurement of the Hubble constant. Here, we use simulated photometry to investigate the dependency of the TRGB luminosity on stellar age and metallicity as a function of wavelength ($\lambda$ 475 nm $-$ 4.5 $\micron$). We find intrinsic variations in the TRGB magnitude to increase from a few hundredths of a magnitude at $\lambda800-900$ nm to $\sim$0.6 mag by $\lambda1.5\micron$. We show that variations at the longer infrared wavelengths can be reduced to 0.02$-$0.05 mag (1$-$2\% accuracy in distance) with careful calibrations that account for changes in age and metal content. These represent the minimum uncertainties; observational uncertainties will be higher. Such calibration efforts may also provide independent constraints of the age and metallicity of stellar halos where TRGB distances are best measured. At 3.6 and 4.5 $\micron$, the TRGB magnitude is predicted to vary by $\sim0.15$ mag even after corrections, making these wavelengths less suitable for precision distances. 

\end{abstract} 

\keywords{galaxies:\ distances and redshifts -- stars:\ Hertzsprung-Russell and CM diagrams -- galaxies:\ halos -- cosmology:\ distance scale}

\section{The Tip of the Red Giant Branch as a Standard Candle}\label{sec:intro}
Distances are one of the fundamental and essential measurements in astronomy, placing a myriad of properties of astronomical systems on absolute scales. Distances to galactic systems have additional roles to play. In the nearby Universe, building a library of accurate, extragalactic distances has helped map the visible structure and flows of galaxies in the Laniakea supercluster of galaxies \citep{Tully2014}. In the larger Universe, extragalactic distances are used to build the cosmic distance ladder, which enables a precision measurement of the Hubble constant \citep[($H_0$, e.g.,][]{Riess2011, Freedman2012, Riess2016, Riess2019} independent of the Planck measurement \citep{Planck2018, Addison2018}. $H_0$ values determined locally and at early times currently disagree at the 4.4$\sigma$ level, suggesting a new feature may be needed in our cosmological model \citep{Riess2019}, and emphasizes the importance of precision distances.

Arguably, the most accurate and inexpensive method for measuring high precision extragalactic distances in the nearby Universe is the tip of the red giant branch (TRGB) method \citep{Mould1986, Freedman1988, DaCosta1990, Lee1993}. The TRGB is a prominent feature in the color-magnitude diagram (CMD), defining the upper boundary of red giant branch (RGB) stars. The TRGB feature arises at the onset of core helium burning in low-mass stars. RGB stars reaching the tip have hydrogen shell-burning supported by an electron-degenerate helium core. When sufficient core temperatures of $\sim10^8$ K are reached, helium burning is initiated, resulting in an abrupt decrease in luminosity and a shift toward bluer colors. This rapid change in brightness creates a discontinuity in CMD space that is readily apparent in optical and near-infrared wavelengths. Because the helium flash depends on the temperature in the electron degenerate stellar core, the TRGB occurs at a predictable luminosity that can be calibrated and used as a standard candle \citep[see e.g.,][and references therein for a recent summary of the TRGB distance method]{Beaton2018}.

\subsection{The TRGB in the Near-Infrared}
TRGB distances are usually measured in the I-band where the TRGB magnitude is independent of total stellar mass (i.e., stellar age) with only a modest dependency on stellar metallicity and reddening correction \citep[e.g.,][]{Lee1993, Salaris1997}. The Hubble Space Telescope (HST) has made it possible to measure TRGB distances to hundreds of galaxies within the Local Volume in the I-band equivalent F814W filter \citep[e.g., CosmicFlows program;][]{Tully2016}. The James Webb Space Telescope (JWST) will enable TRGB distances in the similar F090W filter over even larger volumes due to higher angular resolution and reduced integration times. 

At near-infrared wavelengths redder than the I-band regime, the TRGB becomes brighter as a result of bolometric corrections \citep[for a thorough explanation of bolometric corrections and their impact on TRGB luminosities see][]{Salaris2005}. If the TRGB presents a constant standard candle in the near-infrared that is independent of stellar age and metallicity (or if dependencies on age and metallicity could be accounted for in calibrations), TRGB distances would be possible at reduced observational costs and in a regime that is less affected by extinction \citep[e.g.,][]{Ferraro2000, Bellazzini2004, Dalcanton2012, Serenelli2017, Madore2018, Hoyt2018}. Earlier theoretical work identified a large dependency in the TRGB luminosities on stellar content at near-infrared wavelengths and dismissed these observations as a reliable distance indicator, but the potential calibration of these variations was not discussed \citep{Salaris2005}. 

There have been observational efforts to calibrate the TRGB in the NIR with mixed results. Recent TRGB measurements in the near-infrared JHK bands in the nearby galaxy IC~1613 found that the TRGB magnitudes derived from different regions of the galaxy were consistent, implying a TRGB luminosity independent of any radial gradient in stellar age or metallicity within the galaxy \citep{Hatt2017, Madore2018}. However, the star formation history for this galaxy has been shown to be remarkably homogeneous across similar radii \citep{Skillman2014}. Furthermore, metallicity variations in IC~1613 are expected to be small for two reasons. First, in keeping with its low stellar mass, IC~1613 is metal poor today, so, by definition, the variations in metallicity are limited. Specifically, \citet{Kirby2013} determine the mean stellar [Fe/H] $= -1.19$, with a dispersion of 0.37 dex. Even the current ISM oxygen abundance, as determined from \HII\ regions, is $12+$log(O/H) $= 7.73\pm0.04$ (which translates to [O/H] $= -0.96$ using the solar oxygen abundance from \citet{Asplund2009}) and shows only mild evidence of chemical evolution \citep{Bresolin2007}. Second, the older stellar populations (i.e., TRGB progenitors) of dwarf galaxies appear to be well-mixed radially, resulting in very weak gradients in the metallicity of the older stellar populations \citep[e.g.,][]{Hidalgo2013}. Thus, the TRGB magnitudes at different radii in IC~1613 would be expected to agree as they are measured from stars of similar age and metal content. 

Observations using HST WFC3 imaging with the F110W and F160W filters on nearby dwarfs and spiral galaxies suggest that calibrations may prove challenging due to a potential offset in the absolute calibration of isochrones and the WFC3 IR filters, contamination of non-RGB populations near the TRGB, and, more troubling, the possibility that factors other than age and metallicity, that have yet to be identified, may impact the slope of the RGB on which the calibrations depend \citep{Dalcanton2012}. 

There is some observational evidence that the TRGB does not remain constant across the near-infrared regime. Variability in TRGB magnitudes was reported for a sample of nine nearby dwarf galaxies based on resolved stellar populations imaged at 3.6 $\micron$ with the {\em Spitzer Space Telescope} IRAC instrument \citep{McQuinn2017a}. TRGB distance work is not typically done at 3.6 $\micron$ as contamination from background sources, especially AGN, overlap with and extend past the upper RGB in an IRAC CMD, masking the TRGB at 3.6 $\micron$. To circumvent this, \citet{McQuinn2017a} identified the TRGB at 3.6 $\micron$ {\em in each galaxy} using point sources matched in the HST ACS F814W observations of the same fields of view where AGN and other background contaminants can be eliminated based on their morphology in the higher resolution HST imaging. The TRGB at 3.6 $\micron$ was found to vary by 0.70 mag across the nine galaxies; this range in TRGB magnitude was not clearly correlated with metallicity, suggesting star formation histories or another variable may be important at this wavelength. 

\begin{table}
\begin{center}
\caption{Filters, Stellar Ages, Stellar Metallicities of the Synthetic Photometry}
\label{tab:parameters}
\end{center}
\begin{center}
\vspace{-12pt}
\begin{tabular}{ll}
\hline 
\hline
HST 		& F475W (F606W) \\
 		& F606W (F814W) \\
 		& F814W (F606W) \\
 		& F110W (F160W) \\
 		& F160W (F110W) \\
JWST 	& F090W (F200W) \\
	 	& F115W (F200W) \\
	 	& F150W (F200W) \\
	 	& F200W (F090W) \\
	 	& F277W (F090W) \\
Ground	& J (K) \\
		& H (J) \\
		& K (J) \\
Spitzer	& IRAC [3.6] (IRAC [4.5]) \\
		& IRAC [4.5] (IRAC [3.6]) \\
\\
Age \& Metallicity	& 10 Gyr, $-2.0$ \\
Combinations		&10 Gyr, $-1.5$ \\
				&10 Gyr, $-1.0$\\
				& 5 Gyr, $-1.5$\\
				& 12 Gyr, $-1.5$\\
				& 4 to 12 Gyr, $-2.0$ to $-1.0$\\
				& 0 to 14 Gyr, $-2.0$ to $0.0$\\
\hline \\
\end{tabular}
\end{center}
\vspace{-15pt}
\tablecomments{Synthetic photometry was generated in the above filters, paired with the filters listed parenthetically, from the various observatories for seven different stellar age and isochrone metallicity combinations based on the PARSEC stellar evolution library \citep{Bressan2012} which use the JHK bandpasses from \citet{Bessell1988}.}
\end{table}

\subsection{The TRGB in Stellar Halos}
TRGB distances to massive galaxies (both spirals and ellipticals) are typically measured using stars in the outskirts and halos of galaxies. This is done for a number of practical reasons as there is less confusion from non-RGB populations (i.e., red helium burning (RHeB) stars and asymptotic giant branch (AGB) stars) in the halos of galaxies, there is less extinction, and there is reduced crowding which increases the accuracy of the photometry. 

Importantly, halo populations are also used because of the expectation that age and metallicity variations are small in the halos of present-day galaxies \citep[e.g.,][]{Beaton2016}, which, if true, reduces systematic uncertainties in the TRGB luminosities. However, given the building histories of halos, it is reasonable to expect variations in mean ages and metallicities.

In a comprehensive study of resolved stars in the halos of six spiral galaxies, the GHOST survey \citep{Streich2016} not only found relatively high median metallicities (i.e., [Fe/H] $>$ -1.2 dex) as far out as 70 kpc from the galaxy centers, they also reported metallicity gradients in the halos of half the sample and field-to-field variations in the mean metallicity of RGB stars including differences along the major and minor galaxy axes \citep{Monachesi2016}. The stellar halo of M31 has a strong metallicity gradient of nearly a full dex \citep[e.g.,][]{Ibata2014, Gilbert2014}. Collectively, the median stellar halo metallicity of these spirals vary by nearly a dex over a narrow range in galaxy mass and rotation velocity \citep{Monachesi2016}. Metallicity variations in the stellar halos of elliptical galaxies are less understood, although there is some evidence of metallicity gradients \citep[e.g.,][]{Greene2013}. Thus, assuming stellar halos are consistently metal-poor with little variation does not appear to be valid. To date, few constraints have been placed on the stellar ages.

If precision TRGB distance work is to be extended into the near-infrared where there is potentially a larger dependency on stellar age and metallicity, these dependencies cannot be diminished simply by assuming that stellar halos have similar, consistent, or even well-characterized stellar content. It is quite the opposite. Measuring precision TRGB distances in the near-infrared from stellar halos {\it requires} a full understanding and careful calibration of how the TRGB varies with age and metallicity. Such measurements can then add valuable constraints on the content of stellar halos and the build-up of galaxies. 

\begin{figure*}
\includegraphics[width=\textwidth]{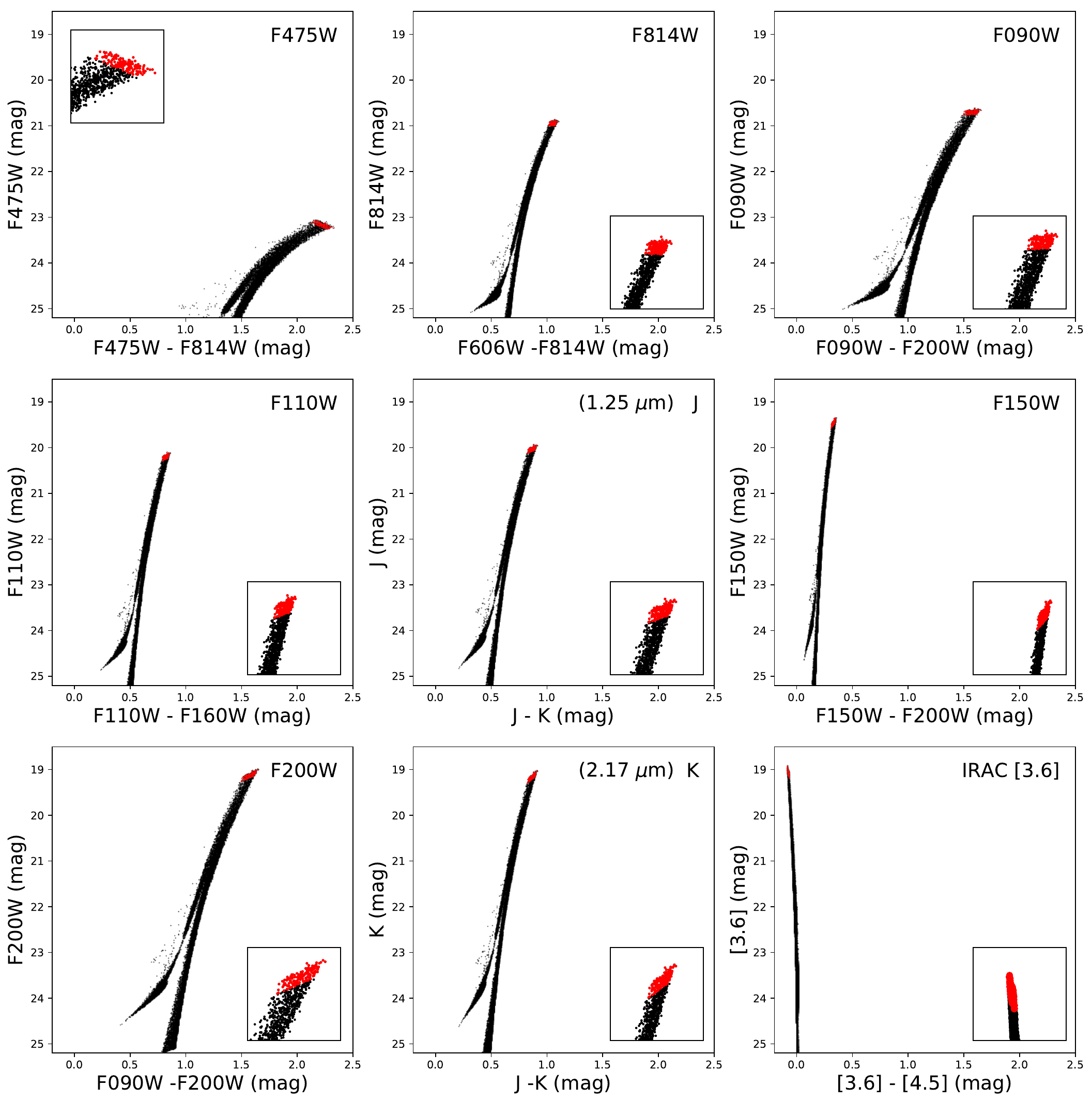}
\caption{Synthetic CMDs for 9 of the 15 filters listed in Table~\ref{tab:parameters} based on a stellar age of 10 Gyr and stellar metallicity of Z$= -1.5$, plotted on the same magnitude and color scales. The stars within 0.1 mag of the TRGB in the F814W are shown in red in each panel. The inset panels show the upper RGB uniformly magnified 3$\times$ to highlight the changing slope of the TRGB: from negative in the blue F475W filter, to positive in the I-band, to nearly vertical in the IRAC [3.6] filter. Note there are no TP-AGB stars above the TRGB as the PARSEC stellar library used does not include stars in this stage of evolution.}
\label{fig:cmds_filters}
\end{figure*}

\vspace{10pt}
The intent of the present work is to use simulations to investigate the degree to which the TRGB is a constant standard candle at wavelengths redward of the I-band over a range of stellar ages and stellar metallicities. Using stellar models, we generate synthetic photometry in a variety of filters spanning from the optical to the near-infrared with populations with ages varying from $0-14$ Gyr and metallicities varying from $-2.0$ to $0.0$ (\S\ref{sec:cmds}). We measure the TRGB using a maximum likelihood technique (\S\ref{sec:measure}) and present a comparison of the TRGB luminosity as a function of wavelength, stellar age, and stellar metallicity (\S\ref{sec:trgb}). We summarize our findings in \S\ref{sec:conclusions}.

\section{Modeling CMDs from the Optical to the Near-Infrared}\label{sec:cmds}
The TRGB distance method requires observations of resolved stellar populations in two bands. While the TRGB magnitude is measured from the discontinuity of the luminosity function in a single filter of interest, the colors of stars determined from two photometric bands are needed to isolate RGB stars in a CMD. Observationally, the colors of the RGB stars are also used as a metallicity indicator to adjust for differences between the metallicity of the galaxy of interest and the metallicity of the stars used in the zeropoint calibration of the TRGB \citep[e.g.,][]{Rizzi2007a, Jang2017}. 

To explore the variation of the TRGB luminosity as a function of wavelength, stellar age, and stellar metallicity, we generated synthetic photometry using the CMD-fitting tool {\sc match} \citep{Dolphin2002a} and the solar-scaled PARSEC stellar evolution library \citep{Bressan2012}. We chose HST optical filters from the ACS and WFC3 instruments, ground-based JHK near-infrared bands, JWST infrared filters on the NIRCam instrument, and the Spitzer Space Telescope IRAC 3.6 and 4.5 $\micron$ bands (similar to the JWST F356W and F444W filters); a list of the filters is provided in Table~\ref{tab:parameters}. While our primary interest is in wavelengths redward of the I-band, we include two HST blue filters for completeness (i.e., F475W and F606W). The synthetic photometry was generated in all filters assuming a randomly populated Kroupa initial mass function \citep[IMF; ][]{Kroupa2001} with no internal or Galactic extinction and no observational uncertainties. We assume a nominal distance of 1 Mpc (distance modulus $= 25$ mag) for the synthetic photometry, placing the apparent magnitudes on a scale consistent with nearby galaxies.

We generated seven different sets of photometry that sample a range in stellar ages and metallicities. We focused primarily on older age, metal-poor stellar populations that are thought to be representative of the stellar populations in the outer disks of spiral galaxies and many low-mass galaxies where TRGB distances are most often measured. We also include one set of synthetic photometry with stars of all ages and metallicities ranging from metal poor (Z$=-2.0$) to metal rich (Z$=0.0$) to illustrate the differences in a CMD with more complex stellar populations.

Specifically, to explore how the TRGB luminosity may be impacted by different stellar metallicities, three sets of photometry assume an instantaneous burst of star formation 10 Gyr ago but have different stellar metallicities values of $-2.0$, $-1.5$, and $-1.0$.\footnote{Note the age of the instantaneous bursts have a spread of log(age) $=$ 0.05 dex and metallicities have a spread of 0.1 dex.} To explore the impact of stellar ages, two additional sets of photometry have a metallicity Z$= -1.5$ with instantaneous bursts of star formation 5 and 12 Gyr ago. The sixth set of photometry was created with constant star formation from 4$-$12 Gyr with metallicities increasing from Z $= -2.0$ to $-1.0$, generally following an age-metallicity relation. Finally, the seventh set has star formation occurring at a constant rate at all times with metallicity increasing from $-2.0$ at early times to 0.0 at the present day. The different combinations of stellar ages and metallicity are listed in Table~\ref{tab:parameters}. 

As low stellar counts can impact the accuracy of a TRGB measurement, the photometry sets were generated multiple times simulating fields of view with different total stellar masses. From experimentation, we found that stellar masses of $5\times10^7$ \msun\ and greater returned consistent TRGB luminosities. For a lower stellar mass of $5\times10^6$ \msun, the upper RGB began to be under-populated and our TRGB measurements were 0.02-0.03 mag fainter than the input TRGB due to the sparseness of stars near the tip. Our results are consistent with previous work on the minimum number of stars needed for precise TRGB distances \citep[e.g.,][]{Makarov2006, McQuinn2013}. 

CMDs were created from the synthetic photometry pairing filters from the same observing facility. For example, the HST F814W filter commonly used for TRGB distance work was paired with the HST F606W filter to generate a F814W vs. F606W-F814W CMD. Similarly, the J-band filter was paired with the K-band to create a J vs. J-K CMD, etc. Figure~\ref{fig:cmds_filters} presents example CMDs for nine filters assuming a 10 Gyr old stellar population with a metallicity of Z$= -1.5$, the fiducial age and metallicity combination assumed to be in the outer regions of spiral galaxies. All CMDs are plotted with the same magnitude and color axis ranges. There is an absence of AGB stars above the TRGB in the CMDs as the PARSEC stellar library used to generate the synthetic photometry does not include thermally-pulsing AGB  (TP-AGB) stars.

There are several features of interest in the CMDs of Figure~\ref{fig:cmds_filters}. First, the discontinuity of the luminosity function for stars approaching the helium flash at the end of the RGB phase of stellar evolution is readily apparent at all wavelengths. Second, the TRGB luminosity increases by 4 magnitudes from blue to near-infrared wavelengths, mainly due to bolometric corrections. The brighter luminosity at longer wavelengths offers an opportunity to reach the required photometric depths for TRGB distances at lower observational expense and to larger distances. Third, the TRGB has a slope which changes from negative in the blue filters to relatively flat at F814W, to positive in redder filters, and, finally, nearly vertical in the IRAC filters. The zoomed inset panels of the TRGBs in Figure~\ref{fig:cmds_filters} highlight these changes. The slope can complicate identifying the discontinuity of the luminosity function in a given filter and, therefore, in practice, the photometry is often slope-corrected to aid in the TRGB identification. This also serves to correct for differences in metallicities between the target galaxy and the calibration galaxies. We discuss this further below. 

\begin{figure}
\includegraphics[width=0.95\columnwidth]{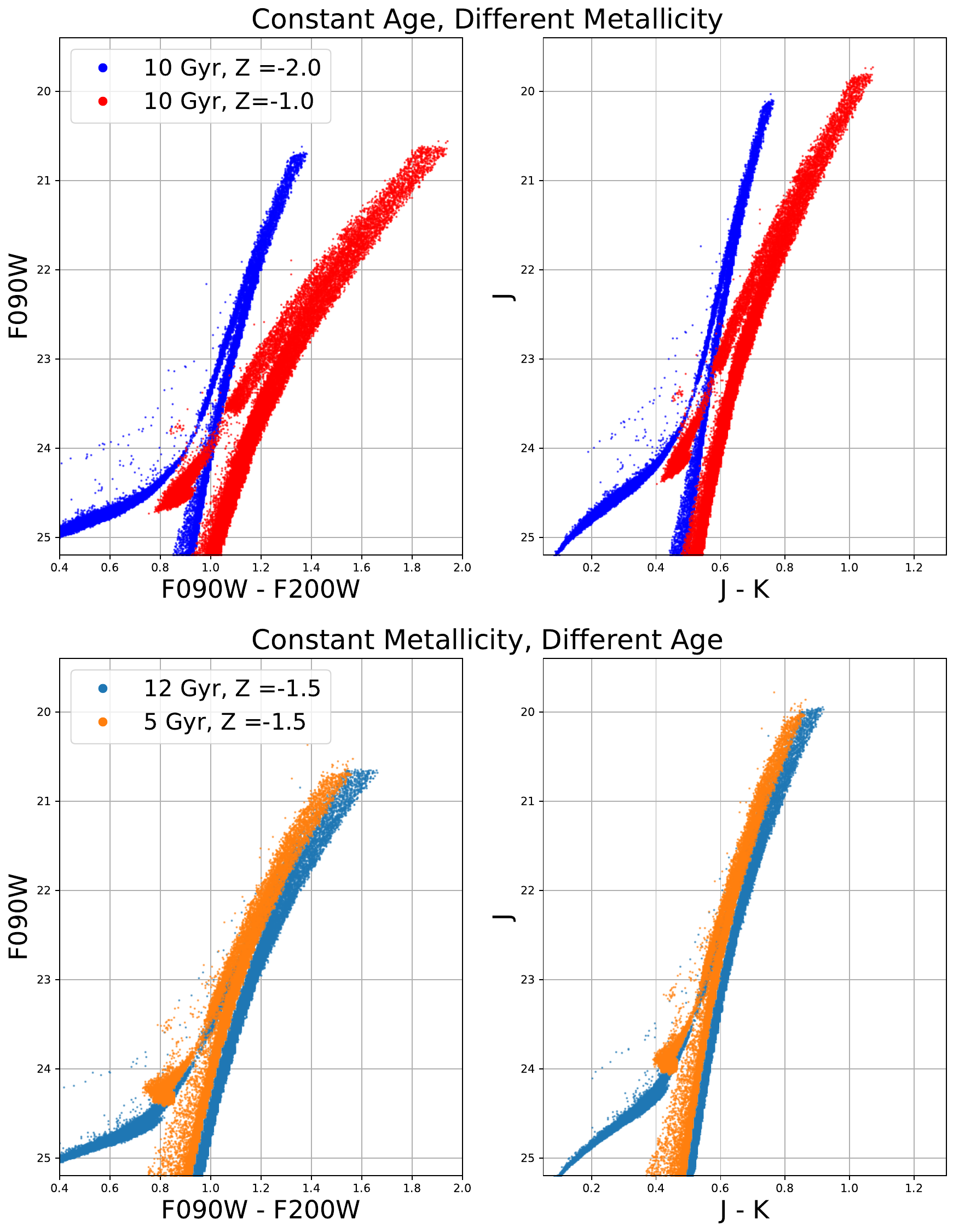}
\caption{CMDs of constant ages with different metallicities (top panels) and of constant metallicity with different ages (bottom panels) for JWST F090W vs. F090W - F200W (left) and J vs. J-K (right) photometry. At 0.9 $\micron$, there is little appreciable change in the TRGB luminosity over the stellar age and metallicities probe; at the longer J-band (1.25 $\micron$), metallicity has a larger impact but differences due to variations in stellar age are also noted. See Figure~\ref{fig:trgb_native} for a quantitative analysis of the differences.}
\label{fig:cmds_age_metal}
\end{figure}

\begin{figure}
\includegraphics[width=0.95\columnwidth]{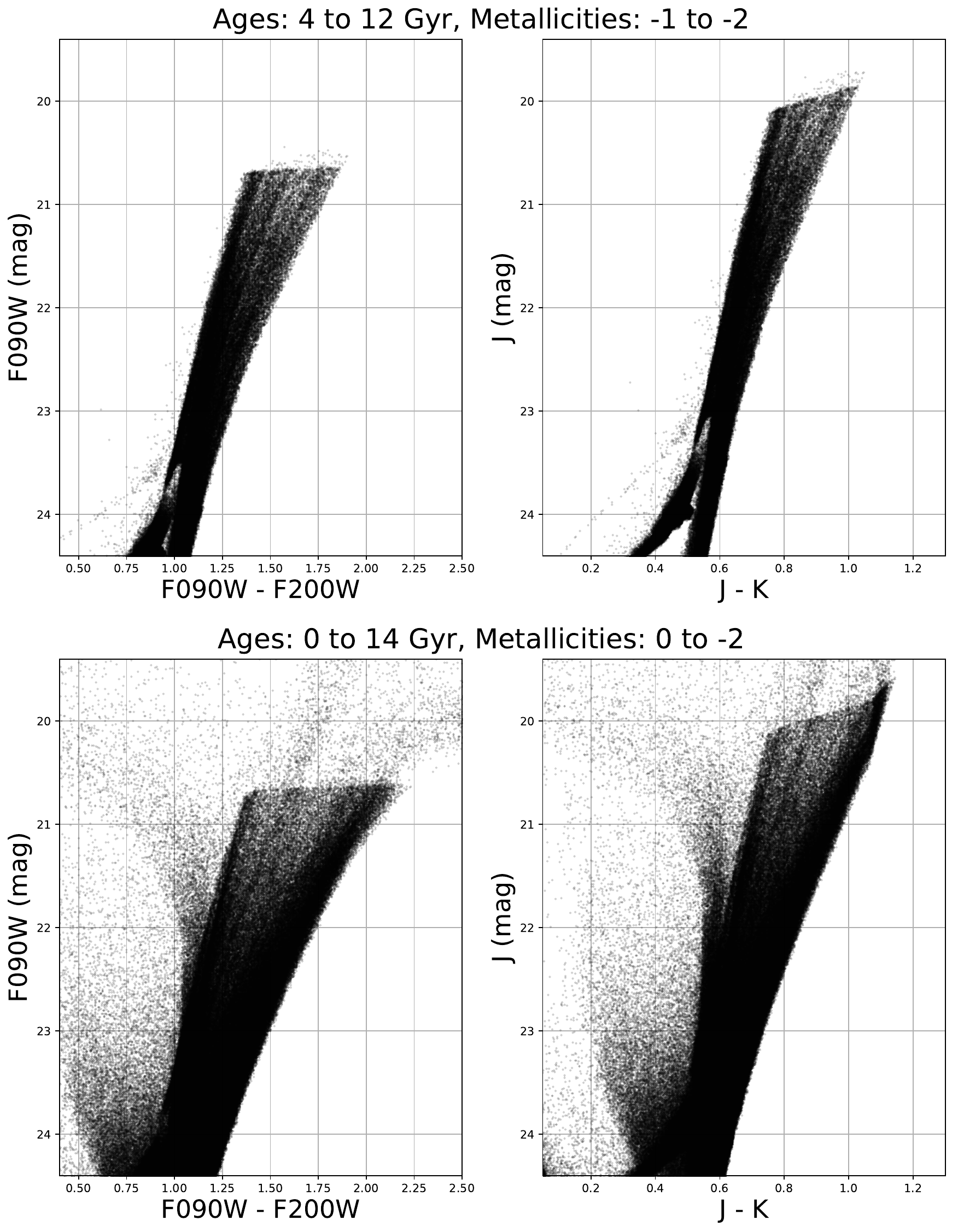}
\caption{CMDs with ages ranging from 4 to 12 Gyr (top panel) and 0$-$14 Gyr (bottom panel) and metallicities ranging from $-1.0$ to $-2.0$ (top panel) and 0.0 to $-2.0$ (bottom panel) for JWST F090W vs. F090W - F200W (left) and J vs. J-K (right) photometry. The range in ages and, in particular, the range in metallicities broadens the RGBs. Stars with younger ages (bottom panels) populate bluer colors as well as brighter magnitudes in the CMD. The CMDs illustrate how stars of younger ages and with a range in metallicities can add complexity to measuring the TRGB in observational data. In addition, the presence of TP-AGB stars in real galaxies exacerbates the complexity.}
\label{fig:cmds_age_metal_highZ}
\end{figure}

Fourth, the color baseline of the stars in each CMD changes depending on how close in wavelength the two filters in the CMD are and, to some extent, how far on the Rayleigh-Jeans tail of the RGB spectrum the filters probe. In our idealized synthetic photometry with no young stars, no TP-AGB stars, no foreground or intrinsic reddening, and no background, the narrow color baseline does not impact our identification of the TRGB, but in observational data, a narrow color baseline can make it difficult to cleanly separate RGB stars from potentially younger stars (such as red helium burning stars and AGB stars) and contamination (such as foreground stars or background galaxies). 

While Figure~\ref{fig:cmds_filters} shows the fiducial photometry from what is typically assumed from stellar populations in the halo of a spiral galaxy, Figure~\ref{fig:cmds_age_metal} illustrates the impact stars of different ages and metallicities have on the TRGB magnitude for two filter combinations: JWST F090W vs.\ F090W-F200W (left panels) and ground-based J vs. J-K (right panels). The top panels compare two sets of photometry that have the same stellar age (10 Gyr burst) but different metallicity (Z $= -2.0$ and $-1.0$). In the F090W filter, there is a small increase in the TRGB luminosity at the higher metallicity, whereas in the redder J-band filter there is a larger increase of order a few tenths of magnitude. The bottom panels compare photometry with a constant metallicity of Z$= -1.5$ but different stellar ages (5 and 12 Gyr bursts). In the F090W band there is no appreciable change in the TRGB luminosity; in the J-band there is a modest but detectable rise in luminosity for the older stars. Note there are also shifts in color of the TRGBs as metal content and, to a lesser degree, stellar age changes. Observationally, such differences in TRGB color are used as a proxy for differences in metallicity and aid in transforming the TRGB luminosity to an absolute scale, as discussed below.

Stellar populations that have a range in age and metallicities can broaden the RGB, and, if younger stars are present, populate both the CMD at bluer colors and at higher luminosities. Figure~\ref{fig:cmds_age_metal_highZ} shows the CMDs with stellar ages ranging from 4$-12$ Gyr and Z$=-1$ to $-2$ (top panels) and from 0$-14$ Gyr and Z$=0.0$ to $-2$ (bottom panels) for the same filter combinations as shown in Figure~\ref{fig:cmds_age_metal} (i.e., JWST F090W vs.\ F090W-F200W; left panels and ground-based J vs. J-K; right panels). These CMDs illustrate the complexity stars with younger ages and larger ranges in metallicity introduce even in idealized, synthetic photometry. Observationally, photometric uncertainties, incompleteness, reddening, and, in particular, the presence of TP-AGB stars will add additional complexity to these CMDs and to identifying the TRGB, emphasizing the advantages of measuring TRGB distances in older, metal-poor populations. 

\section{Measuring the TRGB}\label{sec:measure}
We measured the TRGB brightnesses in all synthetic photometry twice. First, we simply identified the TRGB magnitude in each filter to explore the variability in the absolute TRGB luminosity at a given wavelength across different stellar ages and metallicities. Second, we used the 10 Gyr burst, Z$=-1.5$ CMD as a our fiducial photometry and `calibrated' the other six sets to this TRGB scale. In this way, we can measure how well such a correction can account for the intrinsic luminosity differences in different aged stellar populations with different metal content.

\begin{figure}
\includegraphics[width=\columnwidth]{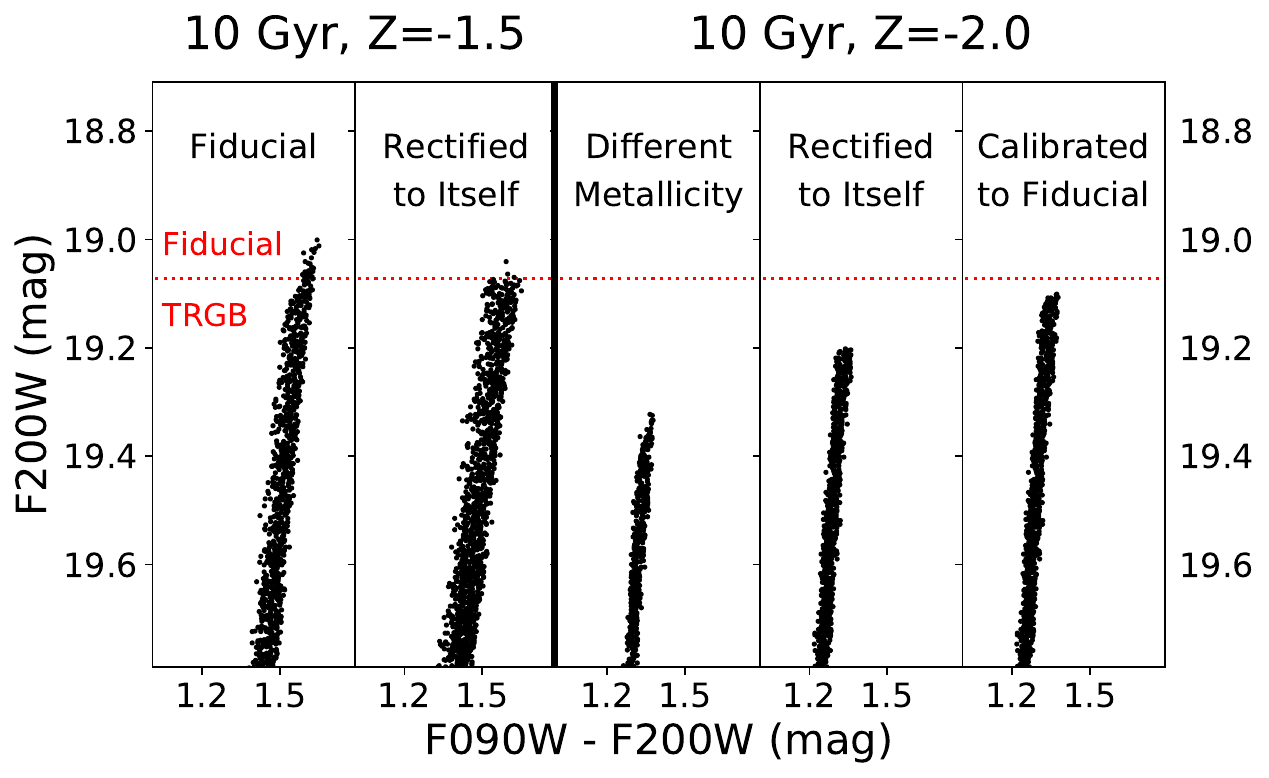}
\caption{{\it Left two panels:} The fiducial photometry is shown in the F090W, F200W filter combination before and after being rectified with the slope correction. {\it Right three panels:} Photometry with a different metallicity is shown that is intrinsically fainter than the fiducial set. After rectifying based on its own slope, the TRGB is flattened, but still remains fainter. After using the slope and color of the fiducial photometry, the TRGB is flattened (allowing for a more robust measurement) and calibrated to nearly the same brightness as the fiducial.}
\label{fig:cmds_transform}
\end{figure}

We first selected stars in the RGB region of the CMDs using simple color and magnitude cuts applied to the photometry. As seen in Figure~\ref{fig:cmds_filters}, the TRGB is sloped to varying degrees, which can complicate the identification of the TRGB discontinuity. Observationally, the RGB is typically `rectified' by transforming the photometry in the target galaxy using the slope and average color of the TRGB in a calibration system. This not only sharpens the discontinuity of the TRGB thereby improving identification of the TRGB \citep[e.g.,][]{Madore2009, McQuinn2016a, McQuinn2016b, McQuinn2017b}, but also acts as a color-based metallicity correction for the zero-point in empirical calibrations \citep[e.g., ][]{Rizzi2007a, Jang2017}.

For our first set of TRGB measurements, we follow a similar approach with the exception that each set of photometry was transformed using its own TRGB slope and average TRGB color. This transformation only rectifies the RGB, without accounting for differences in stellar age or metallicity between different sets of photometry. We demonstrate this in the first four panels of Figure~\ref{fig:cmds_transform} where both the fiducial photometry and a second set of photometry are rectified and the TRGB slope is flattened. We identified stars in the F814W filter within 0.1 mag of the TRGB (i.e., TRGB$+0.1$ mag) and tagged these as upper RGB stars in all filters, shown in red in Figure~\ref{fig:cmds_filters}. The slope of these TRGB stars in each filter was determined using a least squares fit. Each star in each photometry set was then transformed using the following formulism:

\begin{multline}
\label{eq:transform}
MAG_{transformed} = MAG -slope~\cdot \\
[~COLOR - \langle COLOR_{TRGB} \rangle~]
\end{multline}

\noindent Thus transformed, we use a maximum likelihood approach to measure the TRGBs in the synthetic photometry similar to that used in observing programs \citep[e.g.,][]{Makarov2006, McQuinn2014, McQuinn2016a, McQuinn2016b, McQuinn2017b}. 

Our final goal is to determine whether the TRGB can be an effective standard candle at wavelengths redward of the I-band regardless of stellar content. Thus, for our second set of TRGB measurements, we use our fiducial photometry (10 Gyr burst, Z$=-1.5$) as a calibrator and transform the other five photometric data sets to this fiducial system (i.e., applying Equation~\ref{eq:transform} to each photometric set using the slope and average TRGB color from the fiducial system). We demonstrate the effect of this calibration in the final panel of Figure~\ref{fig:cmds_transform} where the 10 Gyr, Z$=-2.0$ photometry is calibrated to the fiducial set. By using the slope and color of the fiducial TRGB, the tip is not only flattened but becomes brighter. We then re-measure the TRGB discontinuity using a maximum likelihood technique. 

\section{The TRGB as a function of Wavelength, Stellar Age, and Metallicity}\label{sec:trgb}
\begin{figure}
\includegraphics[width=\columnwidth]{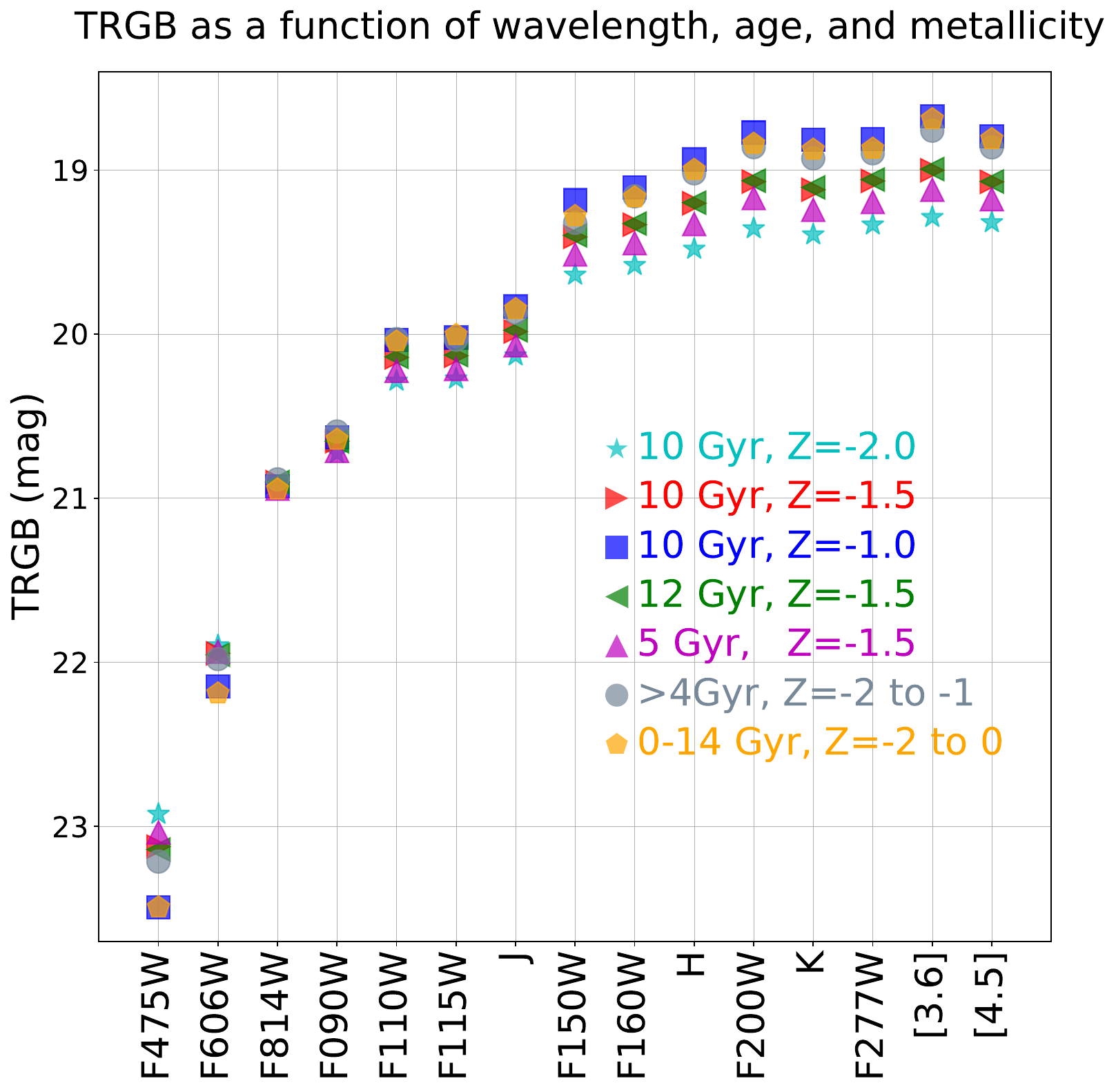}
\caption{Intrinsic TRGB magnitude recovered from the synthetic photometry as a function of wavelength for an assumed distance of 1 Mpc. The different color-symbol combinations show the TRGBs for different stellar age and metallicity combinations.}
\label{fig:trgb_native}
\end{figure}

The TRGB magnitudes as a function of wavelength, stellar age, and metallicity, for a system at a fiducial distance of 1 Mpc, are presented in Figure~\ref{fig:trgb_native}. These TRGB magnitudes are the results of our first fits where the photometry was rectified to itself, without correcting for stellar content. Consistent with the CMDs in Figure~\ref{fig:cmds_filters}, the TRGB luminosity increases in brightness by 4 magnitudes from the optical to the near-infrared. For a given filter, the TRGB luminosities vary due to changes in the age and metallicity of the stars. In the F814W filter, this variability is minimal and the TRGB magnitude is remarkably constant across stellar ages of $4-12$ Gyr and stellar metallicities of $-2.0$ to $-1.0$. The constancy highlights the appropriateness and accuracy of applying the TRGB distance method in the I-band or equivalent filter in nearby galaxies that may have a range in stellar content, confirming what has long been noted observationally that the TRGB in the I-band is an excellent Population II standard candle. The bluer filters (F475W, F606W) show slightly more variability but are significantly fainter than F814W, making these less efficient for TRGB distance work (in addition to requiring larger reddening corrections). We include them for completeness but do not discuss them further. In filters long-ward of 1.5 $\micron$, the intrinsic variability due to stellar age and metallicity content increases to more than 0.6 mag. 

\begin{figure}
\includegraphics[width=\columnwidth]{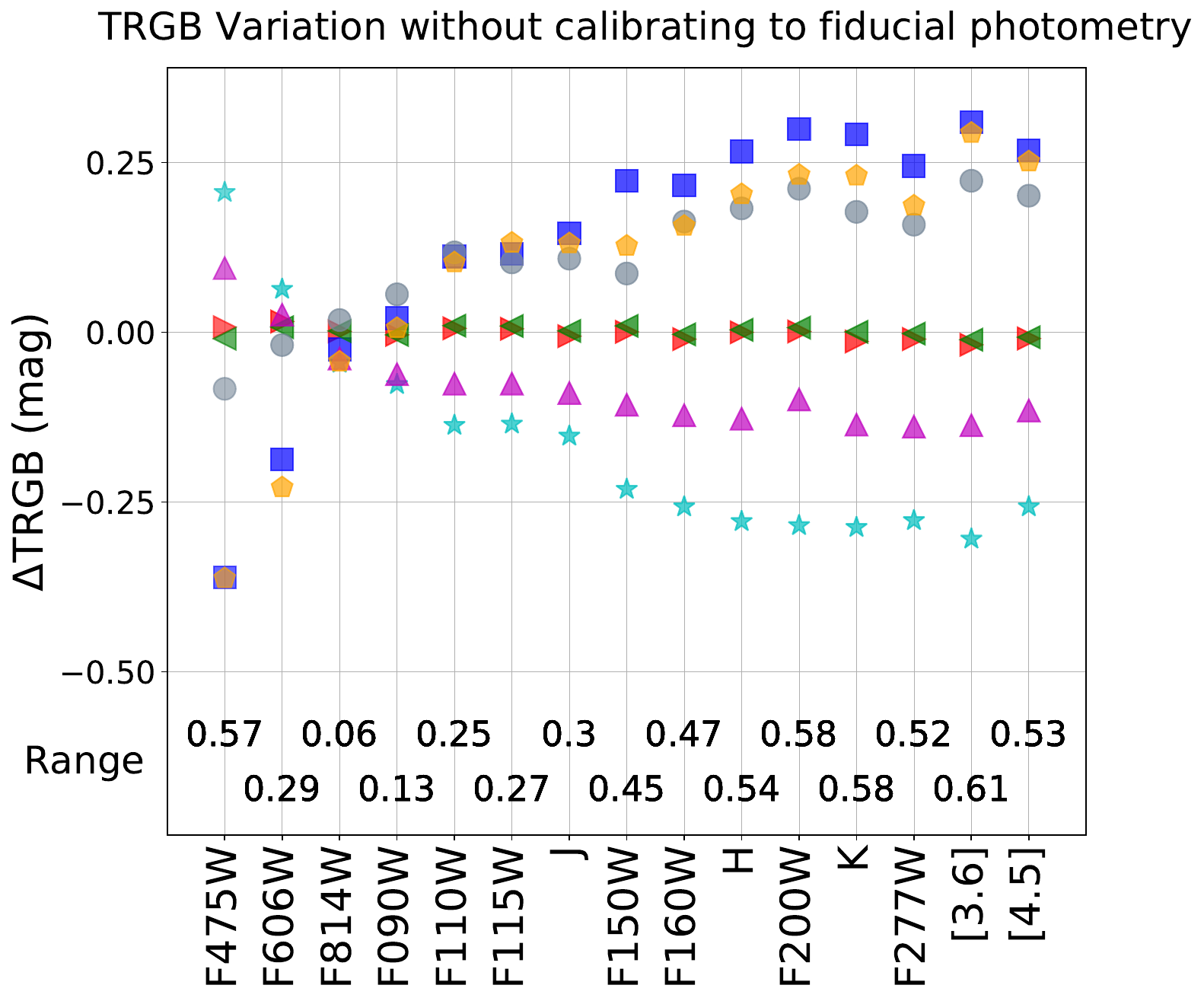}
\caption{Differences in the native TRGB magnitudes from the fiducial photometry as a function of wavelength from Figure~\ref{fig:trgb_native}; the range in TRGB magnitudes for a given filter is listed at the bottom. The more metal poor populations become fainter at longer wavelengths (cyan stars) while the more metal rich populations become brighter (blue squares). Younger populations become fainter for a given metallicity (magenta triangles). However, because metallicity has a stronger impact on the luminosity of the TRGB than age, younger populations that are also more metal rich are brighter than the fiducial photometry (gray circles).  Plot symbols and colors are the same as in Figure~\ref{fig:trgb_native}.}
\label{fig:trgb_untrans}
\end{figure}

We quantify how much the TRGB magnitudes change as a function of stellar content in Figure~\ref{fig:trgb_untrans} where we show the {\it difference} between the TRGB magnitudes relative to the fiducial photometry. Overall, the largest changes in TRGB magnitudes are due to changes in stellar metallicity and these changes increase with increasing wavelength. The metal poor stellar populations (cyan stars) become fainter at longer wavelengths while the more metal rich populations (blue squares) become brighter. Note, however, that relative to the fiducial photometry the impact of metallicity on the luminosity reverses as a function of wavelength: the highest metallicity photometry (yellow pentagons) is fainter (brighter) at shorter (longer) wavelengths due to changes in the bolometric corrections. Changes in the TRGB magnitude are minimized in the F814W filter. 

There are also measurable changes in TRGB magnitudes from changes in stellar age. Focusing on the synthetic photometry at Z $= -1.5$ with ages of 5 and 12 Gyr (magenta triangles; green leftward triangles), the younger stars are fainter redward of the I-band. The synthetic photometry that include a range of stellar ages stellar metallicities (4-12 Gyr, Z $=-2.0$ to $-1.0$ gray circles; 0-14 Gyr, Z $=-2.0$ to $0.0$ yellow pentagons) lie towards the top of the distributions. At longer wavelengths, the higher metallicity stars increase the TRGB luminosity, but because these stars are younger, the TRGB brightness is lower than a stellar population that is metal poor but has only older stars (i.e., 10 Gyr, Z$= -1.0$; blue squares). 

The $\sim0.6$ mag range in absolute TRGB magnitude at longer wavelengths has important implications for designing observing strategies. Metal poor stellar populations will require longer exposure times to reach required photometric depths for accurate TRGB measurements. Younger stellar populations may also require longer exposure times, but this could potentially be offset if the younger populations are more metal-rich, as would be expected in galaxies in the present-day Universe.

\begin{figure}
\includegraphics[width=\columnwidth]{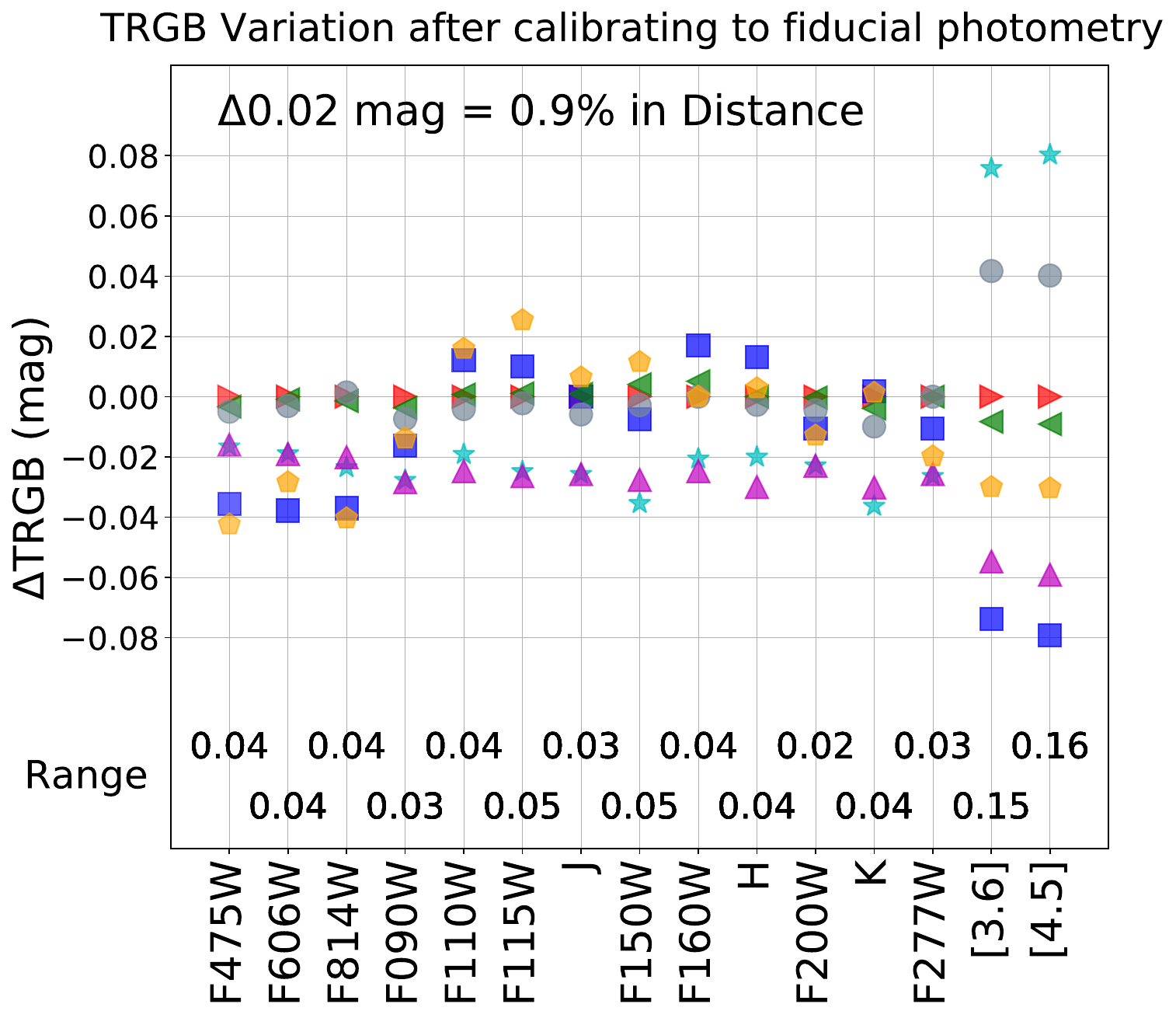}
\caption{Differences in TRGB magnitudes after calibrating to the fiducial photometry (10 Gyr burst, Z$=-1.5$); the range in TRGB magnitudes for a given filter is listed at the bottom. The recovered TRGB magnitudes are remarkable constant for all ages and metallicities modeled with a spread of $0.02-0.4$ mag for all filters except for the IRAC bandpasses. Plot symbols and colors are the same as in Figure~\ref{fig:trgb_native}.}
\label{fig:trgb_trans}
\end{figure}

Once the photometry is calibrated to our fiducial data set (10 Gyr burst, Z$=-1.5$) using the color-based metallicity correction (i.e., our second set of TRGB measurements from \S\ref{sec:measure}), the TRGB magnitudes become remarkably similar across nearly all filters. Figure~\ref{fig:trgb_trans} presents the {\it differences} in TRGB magnitudes between the fiducial photometry and each of the other five sets of photometry for each filter. The variation in TRGB magnitudes is minimized after this correction and, as shown at the bottom of the panel, varies from $0.02-0.04$ mag for filters blue-ward of the IRAC band passes. This remaining $0.02-0.05$ range in TRGB magnitudes is due to differences in the age and metal content of the stellar populations from the fiducial photometry that we are unable to correct for using the color-based correction and corresponds to an uncertainty in distance of $0.9-2.0$\%. The IRAC 3.6 and 4.5 $\micron$ filters show a more significant spread in TRGB magnitudes indicating that these wavelengths are not optimal for high-precision TRGB distances, consistent with observational measurements of the TRGB in the IRAC filters \citep{McQuinn2017a}. 

The 0.9 $-$ 2.0\% precision in distance represent the minimum uncertainties in this wavelength regime based on an idealized case from synthetic photometry. Variations will be larger with observational data. For example, \citet{Dalcanton2012} explored the TRGB in the F110W/F160W filter combination with observations on a sample of 23 nearby galaxies. Their analysis revealed an offset between the color and absolute magnitude of the TRGB with current isochrone models, difficulties in accurately measuring the TRGB luminosity due to the presence of AGB and RHeB stars, and a large dispersion in correlating the slope of the RGB with metallicity. These results highlight that, while there is great potential for the NIR TRGB to be a precise distance indicator, careful calibration with a large empirical data set is needed.

\section{Conclusions}\label{sec:conclusions}
We have measured the TRGB magnitude as a function of wavelength (457 nm - 4.5 $\micron$), stellar age, and stellar metallicity in synthetic photometry generated from the PARSEC stellar library. The primary metallicity (Z $= -2.0$ to $-1.0$) and age (4$-$12 Gyr) range probed by our simulated photometry was chosen to cover the expected stellar properties in the outskirts and halos of massive galaxies and in many low-mass galaxies where TRGB distances are most often measured observationally. We also explored one set of photometry that included stars of all ages over a larger range in metallicity to demonstrate how more complex stellar populations impact a CMD and the TRGB measurement.

In summary, we find:
\begin{itemize}
\item The TRGB becomes brighter by $\sim2$ magnitudes from the HST I-band equivalent F814W filter to the Spitzer IRAC 4.5 $\micron$ filter (comparable to the JWST F444W filter), enabling observational gains if the TRGB magnitude can be carefully calibrated in the near-infrared.

\item In the I-band equivalent HST F814W filter and JWST F090W filter, the TRGB is remarkably constant across all ages and metallicities probed. In the slightly redder HST F110W, JWST F115W, and ground-based J filters, the TRGB magnitudes show variability up to 0.3 (0.09) magnitudes depending on the metallicity (age) of the stellar population. In filters longward of the J-band, the TRGB magnitude varies by $\sim0.6$ ($\sim0.1$) mag depending on the metallicity (age) of the stellar population. The TRGB luminosities in the near-infrared are fainter for younger and/or metal-poor stellar populations. If the younger populations are also more metal-rich, the TRGB will be brighter due the stronger influence of metallicity on the luminosity. Metal-poor populations will be fainter, regardless of stellar age, which has important implications for designing observing strategies of such systems and determining integration times needed to reach sufficient photometric depths for secure TRGB detections.

\item After applying a TRGB slope and color-based correction using a set of fiducial synthetic photometry (i.e., a 10 Gyr stellar population with Z$=-1.5$), the spread in the measured TRGB magnitudes is minimized to $0.02-0.05$ mag in filters blueward of the IRAC bandpasses, corresponding to an uncertainty in distance of 0.9 $-$ 2.0\%. Thus, with careful empirical calibrations that can correct for {\em both stellar age and metallicity}, near-infrared filters offer increased observational efficiency for measuring the TRGB. For example, accurate TRGB distances could be achieved using the JWST F277W filter with a gain of 2 magnitudes in brightness over the I-band. In the absence of such calibrations, TRGB distances in the near-infrared are vulnerable to systematic uncertainties of $\sim$30\% (i.e., $\Delta$dm $\sim0.6$ mag). 

\item We stress that the 0.9$-$2.0\% precision in distance represents the {\em minimum} uncertainties in this wavelength regime. Uncertainties on TRGB magnitudes measured from real observational data will be higher due to photometric uncertainties, the presence of TP-AGB and potentially larger number of RHeB stars, reddening uncertainties, and the accuracy of an empirically determined calibration. 

\item Even after applying a slope and color-based correction using a set of fiducial synthetic photometry, the TRGB magnitudes varied by $\sim0.15$ mag in the IRAC 3.6 and 4.5 $\micron$ band passes (which are similar to the JWST F356W and F444W filters). The increased sensitivity to age and metal content and steep RGB slope makes the TRGB at these redder wavelengths less desirable as a precision distance indicator, consistent with  TRGB magnitudes measured observationally from Spitzer 3.6 $\micron$ data \citep{McQuinn2017a}.

\item As stellar halos appear to have metallicity gradients and mean metallicities that vary by as much as 1 dex between galaxies of similar masses \citep{Monachesi2016}, TRGB investigations in near-infrared wavelengths could be an independent probe of the stellar ages and metallicities in galaxy halos and may provide important insight into the mass build-up of galaxies.
\end{itemize}


\renewcommand\bibname{{References}}
\bibliographystyle{apj}
\bibliography{../../../bibliography.bib}

\end{document}